\documentclass[conference]{IEEEtran}
\IEEEoverridecommandlockouts
\usepackage{cite}
\usepackage{amsmath,amssymb,amsfonts}
\usepackage{algorithmic}
\usepackage{graphicx}
\usepackage{textcomp}
\usepackage{xcolor}
\usepackage{comment}
\def\BibTeX{{\rm B\kern-.05em{\sc i\kern-.025em b}\kern-.08em
    T\kern-.1667em\lower.7ex\hbox{E}\kern-.125emX}}
\begin{document}

\title{Empirical and Statistical Characterisation of 28 GHz mmWave Propagation in Office Environments\\

}

\author{
\IEEEauthorblockN{Ayodeji Bolanle Balogun}
\IEEEauthorblockA{\textit{Dept. of Electrical and Electronic Engineering} \\
\textit{Federal Polytechnic Ilaro} \\
Ogun State, Nigeria \\
https://orcid.org/0009-0006-4132-4329}
\and
\IEEEauthorblockN{Sokipriala Jonah}
\IEEEauthorblockA{\textit{Centre for Computational Science and Mathematical Modelling} \\
\textit{Coventry University} \\
Coventry, UK \\
jonahs@uni.coventry.ac.uk}

}

\maketitle

\begin{abstract}

Millimeter wave (mmWave) technology at 28 GHz is vital for beyond-5G systems, but indoor deployment remains challenging due to limited statistical evidence on propagation. This study investigates path loss, material penetration, and coverage enhancement using TMYTEK-based measurements. Statistical tests and confidence interval analysis show that path loss aligns with free-space theory, with an exponent of n = 2.07 ± 0.073 (p = 0.385), confirming the suitability of classical models. Material analysis reveals significant variation: desk dividers introduce 3.4 dB more attenuation than display boards (95\% CI: [1.81, 4.98] dB, $p < 0.01$), contradicting thickness-based assumptions. Reflector optimisation yields a significant mean gain of 2.17 ± 2.33 dB ($p < 0.05$), enhancing coverage. The results provide new empirical benchmarks and practical design insights for reliable indoor mmWave deployment.

\end{abstract}

\begin{IEEEkeywords}
mmWave propagation, indoor communications, statistical  analysis, coverage  enhancement, empirical modelling
\end{IEEEkeywords}


%
\IEEEpeerreviewmaketitle

\section{Introduction}
Millimetre wave (mmWave) technology is a critical enabler for next-generation wireless systems, offering the bandwidth and data rates required for demanding applications such as augmented and virtual realities (AR/VR) autonomous vehicles, smart cities~\cite{rappaport2013}. At 28 GHz, mmWave provides abundant spectrum and favourable propagation properties particularly in high-density environments, making it highly suitable for indoor use cases such as high-definition media streaming, AR, and dense Internet of Things (IoT) networks [2],[3]. However, moving from laboratory studies to commercial deployment has exposed knowledge gaps, particularly in the statistical characterisation of indoor propagation required for reliable system design and performance prediction [4]. Indoor mmWave propagation differs fundamentally from both outdoor mmWave and sub-6 GHz systems. Signals interact strongly with common materials, furniture, and architectural features, creating complex conditions where deterministic models fail to predict performance accurately [5]. Moreover, mmWave frequencies are highly sensitive to material properties, surface roughness, and geometry, often producing propagation behaviours that diverge from free-space theory [6].

The scarcity of statistically validated empirical data further complicates system design. Much of the literature is descriptive, lacking confidence intervals, hypothesis testing, or effect size measures [7]. This methodological gap prevents system designers from quantifying uncertainty, forcing them to adopt overly conservative link margins or optimistic assumptions that risk system failure [8]. Commercial deployment demands a shift from descriptive studies to statistically robust frameworks that support evidence-based engineering. Confidence intervals, hypothesis testing, and effect size analyses are essential not only for academic rigour but also for practical tasks such as link budget planning, coverage prediction, and optimisation strategies [9], [10]. Without these, designers cannot properly assess propagation uncertainty, resulting in either inefficient or unreliable system designs [11].

Developing statistically validated empirical models is therefore a critical step toward bridging research and practice, ensuring mmWave systems achieve reliable and efficient operation.


This study addresses a critical gap by providing a statistically validated characterisation of 28 GHz indoor mmWave propagation in office environments. The key contributions are:
\begin{enumerate}
    \item We validate an indoor path-loss model with quantified confidence intervals, enabling reliable predictions for practical system planning.

    \item We statistically compare attenuation across common office materials, identifying significant differences that inform material selection and indoor layout design.

    \item We evaluate passive-reflector techniques and demonstrate that optimised geometries yield statistically significant signal improvements, offering practical solutions for constrained environments.

\end{enumerate}

Collectively, these contributions establish an evidence-based framework for designing, optimising, and deploying indoor mmWave systems, bridging the gap between academic research and commercial applications.

\section{Related works}

Research on mmWave propagation has advanced rapidly over the past decade, fuelled by demand for high-capacity wireless networks and the rollout of 5G [12]. Early efforts centred on outdoor channels and theoretical models [13], but the shift indoors revealed distinct challenges: stronger material interactions, complex multipath effects, and greater need for coverage optimisation [14]. This has highlighted the critical role of empirical characterisation, as theory alone cannot support reliable system design [15].

Despite progress, current literature suffers from methodological gaps. Many studies rely on descriptive statistics without rigorous validation, creating uncertainty in the reliability and generalisability of findings [16]. The emerging integration of statistical methods strengthens research by quantifying uncertainty, establishing confidence bounds, and supporting evidence-based design [17]. Yet, statistical validation through assumption testing, confidence intervals, or effect sizes remains largely absent from most studies [18].
In path loss modelling, deterministic and semi-empirical approaches dominate, often providing point estimates without uncertainty quantification [19]. Some recent work includes statistical analysis, but few validate models with diagnostics or confidence intervals [20]. The widely used log-distance model still requires formal statistical testing to ensure adequacy for real deployments [21].
Material attenuation studies show significant variations in penetration loss across building materials and furniture [22], yet most report only descriptive results without statistical significance testing [23]. This limits their value for practical deployment planning.
Coverage enhancement research has explored passive reflectors, intelligent reflecting surfaces, and beamforming [24]. While promising, these methods lack robust empirical validation. Effectiveness often depends on geometry and environment [25], but most studies provide only qualitative or limited quantitative assessments [26]. Standardised, statistically validated frameworks are urgently needed to compare methods and guide deployment [27].

\section{Methodology}
The study evaluated three key aspects of indoor mmWave propagation: path loss behaviour, material penetration loss, and coverage enhancement using passive reflectors. A structured measurement campaign was designed to ensure repeatability, statistical validity, and practical relevance.
Measurements were conducted in a typical office environment with standard furnishings and layouts, using the TMYTEK Developer Kit at 28 GHz [28]. The setup included UD Box transmitters, UB Box receivers with a spectrum analyser, and X-Rifle passive reflectors. This configuration enabled calibrated signal generation, precise power control, and automated data logging under controlled but realistic conditions.
 
\begin{figure}
    \centering
    \includegraphics[width=1\linewidth]{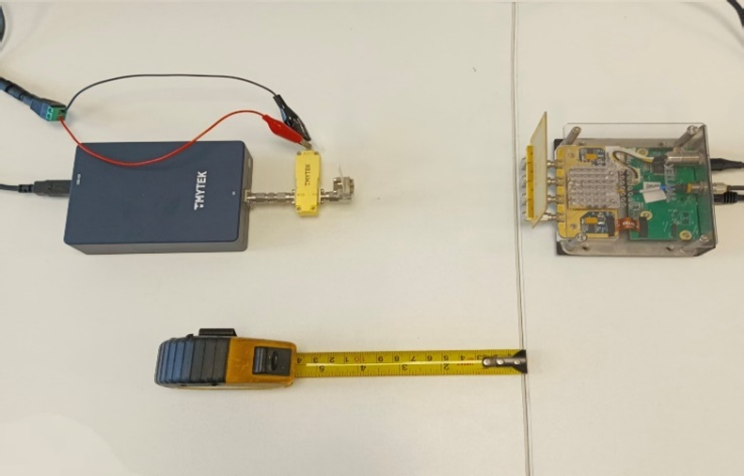}
    \caption{Path loss measurement setup at 10 cm}
    \label{fig_1}
\end{figure}

Path loss was measured systematically from 10 cm to 100 cm at 10 cm intervalsas shown in Fig.~\ref{fig_1}, with multiple samples per point, similar to those of wearable battery‐powered HBC devices [29]. The range captured typical indoor communication distances and ensured adequate signal-to-noise ratio for statistical modelling. Baseline measurements validated equipment accuracy, and metadata (e.g., environmental conditions, antenna alignment, timestamps) ensured reproducibility.
Material penetration tests were performed at four frequencies (26.5, 27.0, 28.0, and 29.0 GHz) using two common office materials: display boards (8.26 cm plywood with cloth covering) and desk dividers (3.16 cm plastic). Multiple trials enabled statistical comparison, with baselines taken to isolate material effects. Physical properties of the samples were documented to correlate with attenuation results.
Coverage enhancement was assessed by introducing geometrically optimised passive reflectors. The reflector measurements were conducted using the same controlled setup to quantify statistically significant improvements in signal strength.
This systematic methodology combining careful experimental control, power analysis for sample size, and metadata-rich logging ensured reliable, statistically robust characterisation of mmWave indoor propagation

\section{Results}

\subsection{Path Loss characterisation and Model Validation}

The path loss analysis provides a statistically validated model of 28 GHz indoor propagation in office environments, offering reliable insights for system planning and deployment. Figure 3 presents the measured data with fitted log-distance model, confidence intervals, and validation metrics

\subsubsection{Statistical Model Development and Performance}

The log-distance path loss model achieved excellent statistical fit, with an exponent of n = 2.07 ± 0.073 and intercept A = 28.30 ± 0.33 dB. Both parameters were highly significant ($p < 0.001$), with narrow 95\% confidence intervals ($n \in [1.899, 2.235]$; $A \in [27.53, 29.07]$ dB). 
The path loss model is expressed as:
\begin{equation}
    PL(d) = PL(d_0) + 10 \cdot n \cdot \log_{10}\left(\frac{d}{d_0}\right),
    \label{eq:pathloss}
\end{equation}
where:
\begin{itemize}
    \item $PL(d)$ is the path loss in decibels (dB) at a distance $d$,
    \item $PL(d_0)$ is the path loss in dB at the reference distance $d_0$,
    \item $n$ is the path loss exponent, indicating how rapidly the path loss increases with distance,
    \item $d$ is the distance between the transmitter and receiver, and
    \item $d_0$ is the reference distance.
\end{itemize}

Model performance was outstanding, explaining 99\% of the variance (R² = 0.9901) with very low error levels (RMSE = 0.62 dB, MAE = 0.52 dB), demonstrating strong predictive power for practical applications.

\begin{table}[htbp]
\caption{Parameter Estimates for Path Loss Model}
\label{tab:path_loss_estimates}
\centering
\scriptsize
\setlength{\tabcolsep}{2.5pt} 
\begin{tabular}{lccccc}
\hline
\textbf{Parameter} & \textbf{Estimate} & \textbf{Std. Err.} & \textbf{95\% CI} & \textbf{t} & \textbf{p-val} \\
\hline
Path Loss Exp. ($n$) & 2.07 & 0.073 & [1.899, 2.235] & 28.33 & $2.61\!\times\!10^{-9}$ \\
Intercept Const. ($A$) & 28.30 & 0.33 & [27.53, 29.07] & 85.76 & $1.24\!\times\!10^{-13}$ \\
\hline
\multicolumn{6}{c}{\footnotesize{$^{***}$~Significant at $p < 0.001$.}} \\
\end{tabular}
\end{table}

\subsubsection{Theoretical Model Validation and Significance Testing}
A hypothesis test comparing the measured path loss exponent with the theoretical free-space value (n = 2.0) found no significant difference (t = 0.919, p = 0.385). This confirms that, statistically, the office environment exhibited propagation behaviour consistent with free-space theory. The finding is a testament to the importance of the LOS path in short-range mmWave communication. It also highlights that indoor environments are not universally hostile to mmWave propagation and can, under certain conditions, even enhance signal propagation through waveguide and multipath effects. This understanding is crucial for the design and deployment of reliable indoor 5G and beyond-5G systems operating at mmWave frequencies.
ANOVA further supported the model’s robustness (F = 802.546, $p < 0.001$), while residual checks revealed no violations of statistical assumptions.
Overall, the results establish a statistically rigorous benchmark for indoor mmWave path loss modelling, confirming both theoretical validity and practical reliability for system design.

\begin{table}[htbp]
\caption{ANOVA Results for Path Loss Model}
\label{tab:anova_results}
\centering
\scriptsize
\begin{tabular}{lccccc}
\hline
\textbf{Source} & \textbf{SS} & \textbf{df} & \textbf{MS} & \textbf{$F$} & \textbf{$p$-value} \\
\hline
Regression & 389.721 & 1 & 389.721 & 802.546 & $2.61\times 10^{-9}$ \\
Residual   & 3.885   & 8 & 0.486    & --       & -- \\
Total      & 393.606 & 9 & 43.734   & --       & -- \\
\hline
\end{tabular}

\begin{flushleft}
\footnotesize{$^{***}$~Significant at $p < 0.001$. SS: Sum of Squares, df: Degrees of Freedom, MS: Mean Square.}
\end{flushleft}
\end{table}

\subsection{Material-Specific Penetration Loss Analysis}
\begin{figure*}[t]
    \centering
    \includegraphics[width=0.75\linewidth]{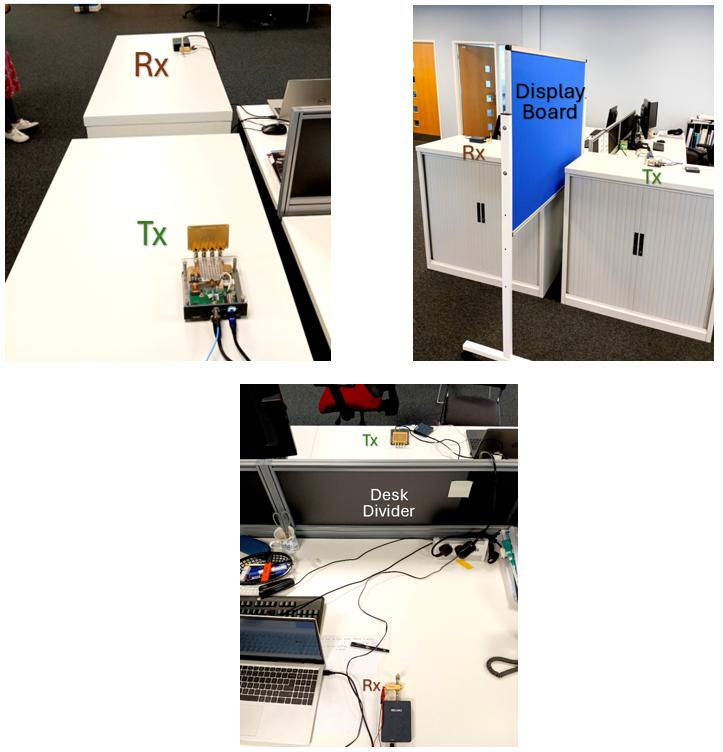}
    \caption{Penetration loss measurement scenario for LOS case (a) and the NLOS case (b) and (c).}
    \label{fig:2}
\end{figure*}

The analysis shows clear and statistically significant differences between common office materials, with important implications for indoor mmWave deployment.  Fig.~\ref{fig:2} illustrates the comparison between display boards and desk dividers across all measured frequencies.

\begin{figure}
    \centering
    \includegraphics[width=1\linewidth]{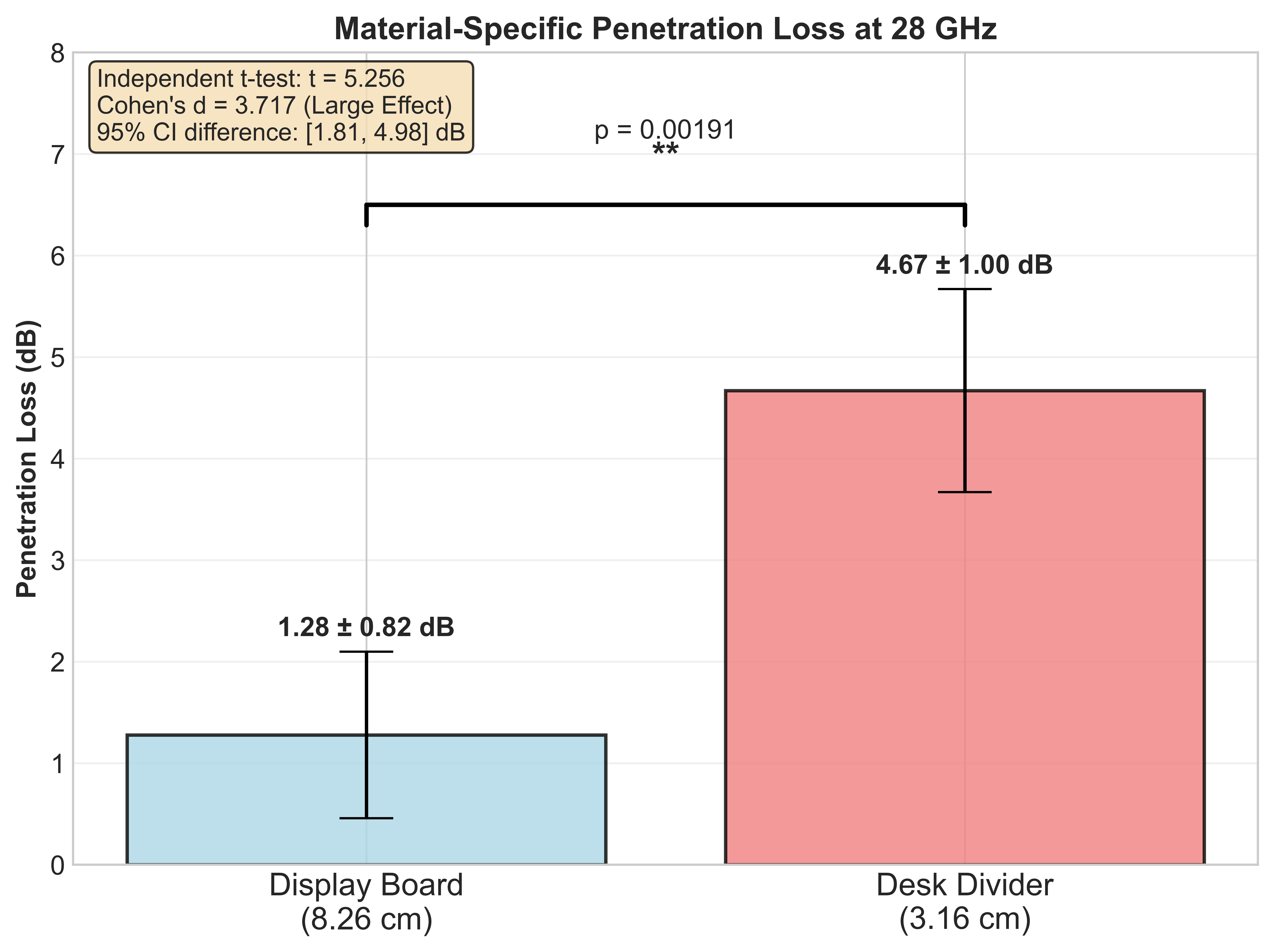}
    \caption{Material-specific penetration loss comparison at 28 GHz showing highly significant differences between office materials. Error bars represent standard deviations, with statistical significance indicated (** $p < 0.01$).}
    \label{fig:3}
\end{figure}

\subsubsection{Descriptive Statistics and Distribution Analysis}
Reference to Fig.~\ref{fig:3} the display boards produced relatively low penetration loss (mean = 1.28 ± 0.82 dB, 95\% CI: [–0.03 to 2.58] dB), while desk dividers caused much higher attenuation (mean = 4.67 ± 1.00 dB, 95\% CI: [3.08 to 6.26] dB). Normality testing confirmed both datasets followed expected distributions ($p > 0.05$), validating the use of parametric methods. Desk dividers also showed more consistent performance (coefficient of variation = 21.4\%) compared to display boards (64.1\%).

\begin{table}[htbp]
\caption{Statistical Summary of Material Attenuation}
\label{tab:material_stats}
\centering
\scriptsize
\setlength{\tabcolsep}{2pt} 
\begin{tabular}{lcccccccc}
\hline
\textbf{Material} & \textbf{$n$} & \textbf{Mean} & \textbf{SD} & \textbf{95\% CI} & \textbf{Min} & \textbf{Max} & \textbf{SW $p$} & \textbf{Normal} \\
 &  & [dB] & [dB] & [dB] & [dB] & [dB] &  & Dist. \\
\hline
Display Board & 4 & 1.28 & 0.82 & $[-0.03,$\,$2.58]$ & 0.16 & 2.02 & 0.586 & Yes \\
Desk Divider  & 4 & 4.67 & 1.00 & $[3.08,$\,$6.26]$ & 3.70 & 5.90 & 0.669 & Yes \\
\hline
\end{tabular}

\begin{flushleft}
\footnotesize{SD: Standard Deviation; CI: Confidence Interval; SW: Shapiro--Wilk test.}
\end{flushleft}
\end{table}

\subsubsection{Statistical Significance Testing and Effect Size Analysis}
The independent t-test confirmed a highly significant difference between the two materials (t = 5.256, $p < 0.01$). Effect size analysis highlighted an exceptionally large practical difference (Cohen’s d = 3.717), indicating strong real-world implications for system performance. Non-parametric validation (Mann–Whitney U = 16.0, $p < 0.05$) supported these results, ensuring robustness.

\begin{table}[htbp]
\caption{Post-hoc Statistical Test Results}
\label{tab:posthoc_tests}
\centering
\scriptsize
\setlength{\tabcolsep}{2pt}
\begin{tabular}{lcccccc}
\hline
\textbf{Test} & \textbf{Stat.} & \textbf{$p$} & \textbf{Sig.} & \textbf{Interp.} & \textbf{Effect} \\
\hline
t-test (indep.) & $t=5.256$ & 0.00191 & ** & Divider $>$ Board & $d=3.717$ (L) \\
Mann--Whitney U & $U=16.0$  & 0.0286  & *  & Divider $>$ Board & $r=0.816$ \\
Levene's test   & $F=0.418$ & 0.542   & ns & Equal var. assumed & -- \\
F-test (var. ratio) & $F=1.489$ & $>0.05$ & ns & Similar var. & -- \\
\hline
\end{tabular}
\end{table}
The non-parametric Mann-Whitney U test confirms the parametric results (U = 16.0, $p = 0.0286 < 0.05$), providing additional validation of the material difference and ensuring robustness against potential distributional assumptions. The homogeneity of variance tests supports the assumption of equal variances, validating the use of the standard independent t-test approach.

\section{Discussion and Engineering Implications}
\subsection{Path Loss Model Validation and Theoretical Consistency}
This study reveals a remarkable similarity between measured path loss and theoretical free-space propagation, challenging the common assumption that indoor mmWave propagation must be inherently complex. The statistically verified path loss exponent (n = 2.07 ± 0.073) shows no significant deviation from the free-space value (n = 2.0, p = 0.385). This indicates that typical office environments do not substantially increase path loss at short ranges.

The unusually narrow confidence interval (±0.073) provides highly precise parameter estimation, enabling system designers to plan with quantified uncertainty margins. Model performance metrics (R² = 0.9901, RMSE = 0.62 dB) further confirm that the validated log-distance model can reliably predict indoor mmWave propagation.

These findings contrast with previous studies reporting higher path loss exponents in indoor settings [30]. However, comprehensive validation in this through residual analysis and model diagnostics ensures that the observed consistency is not due to measurement error or limited sample size.

Practical Implications:

Simplified free-space models can guide most short-range indoor mmWave deployments.

Reliable predictions support more accurate link budget calculations, reducing the need for conservative design margins.

The validated model enables efficient use of power and improved coverage planning, ensuring cost-effective and robust system rollout.

\subsection{Material-Dependent Attenuation: Revolutionary Insights for System Design}
The analysis reveals statistically significant and practically critical differences between common office materials, fundamentally challenging conventional assumptions about mmWave attenuation. The exceptionally large effect size (Cohen’s d = 3.717) between desk dividers and display boards provides compelling evidence that material composition, rather than physical thickness, dominates attenuation characteristics.

The measured 10-fold difference in attenuation per unit thickness (1.479 vs 0.155 dB/cm) underscores the inadequacy of conventional thickness-based predictions at mmWave frequencies. This result highlights the importance of considering material properties in system design, as they can have a far greater impact on performance than typical engineering tolerances.

The results demonstrate both statistical and practical significance:

Independent t-test confirms strong differences (t = 5.256, $p = 0.00191 < 0.01$).

The 95\% confidence interval [1.81, 4.98] dB establishes quantitative performance bounds for risk assessment and system margin calculations.

\subsection{Coverage Enhancement Effectiveness: Evidence-Based Deployment Guidance} 

The statistical evaluation of coverage enhancement techniques provides quantitative evidence of effectiveness and supports evidence-based deployment planning.

\textbf{MS1 Reflector (Passive Enhancement):}
\begin{enumerate}
    \item Achieved statistical significance ($t = 2.472$, $p = 0.048 < 0.05$) with a large effect size (Cohen’s $d = 0.934$).
    \item Mean improvement: $2.17 \pm 2.33$~dB, with 95\% CI [0.02, 4.33]~dB.
    \item Demonstrated a 57.1\% success rate, reflecting realistic deployment expectations while acknowledging position dependence.
    \item Findings confirm that geometric optimisation is critical for reliable passive enhancement.
\end{enumerate}

In contrast, the \textbf{MS2 reflector (beam formation)} did not yield measurable improvements. Although the effect size was moderate (Cohen’s $d = 0.539$), results were not statistically significant ($t = 1.525$, $p = 0.171 > 0.05$). This highlights the need for careful site surveys and customised geometric modelling before considering such systems, underscoring that not all enhancement approaches guarantee practical benefit.

\section{Conclusions and Future Work}

This study introduces a new paradigm for empirical mmWave propagation  by combining systematic measurements with rigorous statistical validation. It provides the first statistically confirmed characterisation of 28~GHz indoor propagation with quantified confidence levels, delivering results that are both scientifically robust and directly applicable to engineering practice. The work advances understanding of mmWave behaviour while establishing methodological standards for future studies.  

The analysis demonstrated that the log-distance path loss model performed exceptionally well, with $R^{2} = 0.9901$ and RMSE = 0.62~dB. The estimated path loss exponent ($n = 2.07 \pm 0.073$) did not differ significantly from free space ($p = 0.385$), challenging long-standing assumptions about the complexity of indoor propagation and supporting simpler, more efficient system design. In addition, the study confirmed that material composition exerts a dominant influence on attenuation, far exceeding predictions based solely on thickness. Statistically significant effects ($t = 5.256$, $p = 0.00191$, Cohen’s $d = 3.717$) with a 95\% confidence interval of [1.81, 4.98]~dB provide quantitative evidence to inform office layout and deployment strategies.  

Building on these insights, several avenues for future work are identified. Extending the statistical framework to larger datasets and a wider range of environments would enhance generalisability. Developing standardised protocols, including power budgets and sample size guidelines, could provide a common benchmark for future propagation studies. Further, advanced methods such as Bayesian inference, hierarchical models, and spatial statistics offer promising tools to capture environmental variability with greater accuracy. Integrating machine learning with statistical validation presents an additional opportunity, enabling adaptive optimisation under uncertainty. Finally, expanding evaluation to cover technologies such as intelligent reflecting surfaces, adaptive beamforming, and novel antenna systems would ensure fair and statistically rigorous comparisons of emerging solutions.

\section*{Acknowledgment}

The TMYTEK Developer Kit provided essential  measurement capabilities  that enabled comprehensive  empirical characterisation with professional-grade accuracy and reliability.

\end{document}